\documentclass[11pt]{article}
\usepackage{amsmath,amssymb,graphicx}
\usepackage{algorithm}
\usepackage{algorithmic}
\usepackage{subfigure}
\usepackage{enumitem}
\usepackage{color}
\usepackage{multirow}
\usepackage[margin=0.8in]{geometry}
\usepackage{setspace}
\usepackage{color}
\usepackage[utf8]{inputenc} 
\usepackage[T1]{fontenc}    
\usepackage{hyperref}       
\usepackage{url}            
\usepackage{booktabs}       

\usepackage{epsfig}
\usepackage{graphicx}
\usepackage{amsmath}
\usepackage{amssymb}
\usepackage{subfigure}
\usepackage{multirow}
\usepackage{array}
\usepackage{algorithm}
\usepackage{algorithmic}
\usepackage[utf8]{inputenc}

\graphicspath{{./figures/}}

\title{Class-Specific Blind Deconvolutional Phase Retrieval Under a Generative Prior}

\author{Fahad Shamshad and Ali Ahmed \\ \small{Department of Electrical Engineering, Information Technology University, Lahore, Pakistan.} \\ \tt{\small{$\lbrace$fahad.shamshad, ali.ahmed$\rbrace$@itu.edu.pk }} }

\begin{document}

\maketitle
\begin{abstract}
In this paper, we consider the highly ill-posed problem of jointly recovering two real-valued signals from the phaseless measurements of their circular convolution. The problem arises in various imaging modalities such as Fourier ptychography, X-ray crystallography, and in visible light communication. We propose to solve this inverse problem using  alternating gradient descent algorithm under two  pre-trained deep generative networks as priors; one is trained on sharp images and the other on blur kernels. The proposed recovery algorithm strives to find a sharp image and a blur kernel in the range of the respective pre-generators that \textit{best} explain the forward measurement model. In doing so, we are able to reconstruct quality image estimates. Moreover, the numerics show that the proposed approach performs well on the challenging measurement models that reflect the physically realizable imaging systems and is also robust to noise. 
\end{abstract}
\section{Introduction}
In this paper, we aim to recover two unknown real-valued signals from the phaseless measurements of their circular convolution. We consider the first signal to be a member of a class of sharp images and second to be a member of a class of blur dataset. Specifically, we want to recover image $\boldsymbol{i} \in \mathbb{R}^{n}$ and blur kernel $\boldsymbol{k} \in \mathbb{R}^{n}$ from the observations of the form
\begin{equation} \label{eq:bd_phase}
\boldsymbol{y} = \vert {\mathcal{A}} (\boldsymbol{i} \circledast \boldsymbol{k}) \vert+ \boldsymbol{n},
\end{equation}
where $\boldsymbol{y} \in \mathbb{R}^{m}$ are phaseless blurry measurements, $\mathcal{A}:\mathbb{R}^{n} \rightarrow \mathbb{C}^{m}$ is the forward operator, and $\boldsymbol{n} \in \mathbb{R}^{m}$ denotes noise perturbation.
The problem is often encountered in numerous imaging applications including visible light communication, Fourier ptychography \cite{holloway2016toward}, and X-ray crystallography where it is easy to build detectors that can measure intensity while discarding the phase information \cite{ahmed2018blind}. The problem \eqref{eq:bd_phase} is notoriously challenging and hard to solve due to its non-linear and non convex nature and without any prior information about $\boldsymbol{i}$ and $\boldsymbol{k}$.

A straight forward approach to restore $\boldsymbol{i}$ and $\boldsymbol{k}$ from phaseless and blurry observations $\boldsymbol{y}$ is to solve the two problems sequentially i.e. solving phase retrieval problem for getting rid of diffraction artifacts, followed by a deblurring algorithm to get estimate of the clean image. However, the simple concatenation of the two models sequentially is sub-optimal due to the error propagation, i.e., the estimated error of the phase retrieval step will be propagated and magnified in the recovery algorithm for blind image deblurring. Moreover, the image reconstruction step is performed twice (one during phase retrieval and other during the deblurring step) that results in increase in the computational cost. This leads to limited applicability of the two-step algorithm in resource-constrained applications. A successful solution to this problem should jointly reconstruct an estimate of the true image and blur kernel from the phaseless blurry and possibly noisy measurements.

Further, in many imaging applications, the ground truth (GT) image is known to belong to a specific class of images such as face or digits etc. This knowledge should be exploited by the image restoration method as general priors learned from an arbitrary collection of natural images are not necessary well-suited for all image classes and often lead to deterioration in performance. Recently, class-specific image restoration methods are gaining relevance and have been shown to be beneficial for many image enhancement applications including deblurring \cite{anwar2018image}, denoising \cite{remez2018class}, super-resolution \cite{li2009example} etc. These class-specific priors have been shown to outperform blanker prior-based approaches.

Inspired from the effectiveness of class-specific priors, in this work, we assume that $\boldsymbol{i}$ and $\boldsymbol{k}$ are not completely arbitrary but are members of some structured classes such as face/digits dataset and motion/Gaussian blurs, respectively. Such structured classes can often be mapped to small vectors lying in a low-dimensional feature space using some unknown (possibly non-linear) generator maps. Our general strategy is to learn the generator maps for each of the class using powerful deep generative networks like generative adversarial network (GAN) \cite{goodfellow2014generative} or variational autoencoder (VAE) \cite{kingma2013auto}. These generator maps introduce very pertinent constraints in \eqref{eq:bd_phase} leading to a more well-conditioned problem. We then use an alternating gradient descent scheme on the latent low-dimensional feature vectors of each of the unknowns (image and blur) to minimize the measurement misfit. We show that this strategy reliably recovers estimates of the true image $\boldsymbol{i}$, and blur kernel $\boldsymbol{k}$. For illustration, we present the results of our proposed algorithm via numerical simulations. 


\textbf{Our Contributions}: In this work, we propose an alternating gradient descent scheme assisted with pre-trained generative priors is able to recover the visually appealing and sharp approaximation of the true image $\boldsymbol{i}$ and blur kernel $\boldsymbol{k}$ from the phaseless blurry measurements $\boldsymbol{y}$. We dubbed our proposed approach as Deep PBD. Specifically, we make the following contributions
\begin{enumerate}[leftmargin=12pt,nolistsep]
\item To the best of our knowledge, this is the first work that aims to handle the challenging problem of phaseless blind image deblurring via leveraging the power of \textit{learned priors} by simply using gradient descent (rather computationally demanding convex optimization based approach as in \cite{ahmed2018blind}). 
\item Further, unlike end-to-end deep learning approaches, Deep PBD is flexible enough to handle variety of forward operators ($\mathcal{A}$), including Fourier and subsampled Fourier ptychography (FP) \cite{holloway2016toward, shamshad2019deep}. 
\item We demonstrate experimentally that Deep PBD is highly robust to noise.
\end{enumerate}

The rest of the paper is organized as follow. In Section \ref{sec:related_work}, we give brief overview of related work. We formulate the problem and give alternating gradient descent based solution in Section \ref{sec:problem_formulation} alongwith description of forward models. Section \ref{sec:experiments} contains experimental results followed by concluding remarks and future directions in Section \ref{sec:conclusion}.





\section{Related Work:} \label{sec:related_work}
Our work is inspired by the two recent works of blind deconvolutional phase retrieval \cite{ahmed2018blind} and blind image deblurring \cite{asim2018solving}. In \cite{ahmed2018blind}, authors aim to solve the phaseless blind image deblurring problem via convex program with rigorous theoretical guarantees. Specifically, by assuming that $\boldsymbol{i}$ and $\boldsymbol{k}$ belong to known random subspaces, the authors resolve the bilinear ambiguity by lifting the phaseless blind deblurring (PBD) problem to higher dimensional space. However, as mentioned in \cite{ahmed2018blind} that their proposed algorithm is less effective for deterministic subspaces that are often encountered in practical applications. One other drawback of their, otherwise appealing, convex program is the computational complexity that is rather high for large scale data and/or for applications where computation time is of the essence. Recently, the author extended their work in \cite{ahmed2019simultaneous},  by utilizing recent advancement in Burer-Monteiro-type approaches \cite{boumal2016non} and perform the optimization in a factored space by solving a series of non-convex programs. Their modified proof guarantee recovery in the presence of noise. 
In \cite{li2019simultaneous}, the authors consider slight variation of problem \eqref{eq:bd_phase} of the form $\boldsymbol{y} = \vert \boldsymbol{Fi} \vert^{2} \circledast \boldsymbol{k} $, where $\boldsymbol{F}$ denotes discrete Fourier transform (DFT) matrix. This model may stem from phase imaging applications that employ partially coherent illumination (e.g.light bulbs, LEDs, and X-ray tubes) \cite{pfeiffer2006phase}. Instead of computationally prohibitive nuclear norm, the authors in \cite{li2019simultaneous} extend the iterative hard thresholding (IHT) algorithm that is widely used in compressive sensing \cite{blumensath2009iterative} and low-rank matrix recovery \cite{tanner2013normalized} to the tensor setting and introduce the tensor IHT (TIHT) algorithm, although in noiseless case.

In \cite{asim2018solving}, authors show the effectiveness of pre-trained generative models for handling the problem of blind image deblurring. Recently, pre-trained generative models have also shown remarkable performance for solving other inverse imaging problems including compressed sensing \cite{bora2017compressed}, Fourier ptychography \cite{shamshad2018robust,shamshad2019adaptive}, phase retrieval \cite{hand2018phase} etc. These pre-trained generative priors bridge the gap between deep learning based approaches (that can take advantage of the powerful learned priors) and conventional \textit{hand designed} priors such as sparsity (that are flexible enough to handle variety of model parameters).
Inspired from their success in inverse imaging problems, we choose to leverage the power of pre-trained generative priors to solve the highly ill-posed problem of phaseless blind image deblurring.

\section{Problem Formulation and Proposed Solution} \label{sec:problem_formulation}

\begin{algorithm}[t]\label{alg:AltGradDescent}
	\caption{Phaseless Deblurring under Generative Priors}
	\begin{algorithmic} 
		\STATE \textbf{Input:} $y$, $\mathcal{G}_\mathcal{I}$, $\mathcal{G}_\mathcal{K}$, and $\mathcal{A}$\\
		\textbf{Output:}  Estimates $\hat{i}$ and $\hat{k}$ \\
		\STATE \textbf{Initialize:}\\
		$\boldsymbol{z}_i^{(0)} \sim  \mathbb{N}(0,1), ~~ \boldsymbol{z}_k^{(0)} \sim  \mathbb{N}(0,1)$ 
		\FOR{${t = 0,1,2,\ldots, T }$}
		\STATE{$ \boldsymbol{z}_i^{(t+1)} \leftarrow \boldsymbol{z}_i^{(t)}$ - $\eta \nabla_{\boldsymbol{z}_i} \mathcal{L}(\boldsymbol{z}_i^{(t)}, z_k^{(t)})$; } \hspace*{\fill} (\ref{eq:Optimization-latent})\\
		\STATE{$ \boldsymbol{z}_k^{(t+1)} \leftarrow \boldsymbol{z}_k^{(t)}$ - $\eta \nabla_{\boldsymbol{z}_k} \mathcal{L}(\boldsymbol{z}_i^{(t)}, \boldsymbol{z}_k^{(t)})$; } \hspace*{\fill} (\ref{eq:Optimization-latent})
		\ENDFOR \\
		$\hat{\boldsymbol{i}}  \leftarrow \mathcal{G}_\mathcal{I}(\boldsymbol{z}^{(T)}_i), \ \hat{\boldsymbol{k}}  \leftarrow \mathcal{G}_\mathcal{K}(\boldsymbol{z}^{(T)}_k)$
   \end{algorithmic}
       \label{alg:generative-prior-deblurring}
\end{algorithm}

We take $\boldsymbol{i} \in \mathbb{R}^n$, and $\boldsymbol{k} \in \mathbb{R}^n$ to be members of some structured classes $\mathcal{I}$, and $\mathcal{K}$, respectively. That is every $\boldsymbol{i} \in \mathcal{I}$, and $\boldsymbol{k} \in \mathcal{K}$ can be characterized by a latent low-dimensional feature vectors $\boldsymbol{z}_i \in \mathbb{R}^o$, and $\boldsymbol{z}_k \in \mathbb{R}^l$, respectively, where $o,l \ll n$. Mathematically, we can write  
$\boldsymbol{i} = \mathcal{G}_\mathcal{I}(\boldsymbol{z}_i),$ $\text{and}$ $\boldsymbol{k} = \mathcal{G}_\mathcal{K}(\boldsymbol{z}_k)$, where $\mathcal{G}_\mathcal{I}: \mathbb{R}^o \rightarrow \mathbb{R}^n$, and $\mathcal{G}_\mathcal{K}: \mathbb{R}^l \rightarrow \mathbb{R}^n$ denotes feature maps.
These feature maps are discovered using generative models trained to learn the probability distributions $p_{\mathcal{I}}$, and $p_{\mathcal{K}}$ of the class $\mathcal{I}$, and $\mathcal{K}$, respectively. This is accomplished by training the  GAN \cite{goodfellow2014generative} or VAE \cite{kingma2013auto} by using representative training data $\{\boldsymbol{i}_q \in \mathcal{I}\}_{q=1}^Q\quad\text{and} \quad \{\boldsymbol{k}_r \in \mathcal{K}\}_{r=1}^R$ of the classes $\mathcal{I}$, and $\mathcal{K}$, respectively.  
After training, we fix the weights of these generative models (pre-trained).
To recover the estimate of the true image $\hat{\boldsymbol{i}}$ and blur kernel $\hat{\boldsymbol{k}}$  from phaseless blurry measurements $\boldsymbol{y}$, we propose minimizing the following objective function in the lower dimensional, latent representation space via gradient descent algorithm

\begin{equation}\label{eq:Optimization-latent}
(\hat{\boldsymbol{z}}_i, \hat{\boldsymbol{z}}_k) := \underset{\boldsymbol{z}_i , \boldsymbol{z}_k}{\text{argmin}}\ \mathcal{L}(\boldsymbol{z}_i,\boldsymbol{z}_k) := \| \boldsymbol{y} - \vert \mathcal{A} (\mathcal{G}_{\mathcal{I}}(\boldsymbol{z}_i) \circledast \mathcal{G}_\mathcal{K}(\boldsymbol{z}_k)) \vert \|^2 \notag \\
 + \gamma\| \boldsymbol{z}_i \|^2 + \lambda\| \boldsymbol{z}_k \|^2, 
\end{equation}
where $\lambda$, and $\gamma$ are free scalar parameters. For brevity, we denote the objective function above by $\mathcal{L}(\boldsymbol{z}_i,\boldsymbol{z}_k)$.  The gradients of the objective w.r.t. $\boldsymbol{z}_i$ and $\boldsymbol{z}_k$ can be easily computed by back propagating through pretrained (known weights) generators. This optimization program can be thought of as tweaking the latent representation vectors $\boldsymbol{z}_i$ and $\boldsymbol{z}_k$, (input to the generators $\mathcal{G}_{\mathcal{I}}$ and $G_{\mathcal{K}}$, respectively)  until these generators generate an image $\mathcal{G}_{\mathcal{I}}$ and blur kernel $\mathcal{G}_{\mathcal{K}}$, respectively, whose circular convolution comes as close to $\boldsymbol{y}$ as possible. The estimated image and the blur kernel are acquired by a forward pass of the latent vectors $\hat{\boldsymbol{z}}_i$ and $\hat{\boldsymbol{z}}_k$ through the generators $\mathcal{G}_{\mathcal{I}}$ and $\mathcal{G}_{\mathcal{K}}$, respectively. Mathematically, $(\hat{i},\hat{k}) = ( \mathcal{G}_\mathcal{I}(\hat{\boldsymbol{z}}_i), \mathcal{G}_\mathcal{K}(\hat{\boldsymbol{z}}_k))$.  
\begin{figure}[t]
\centering
\subfigure{\includegraphics[trim={0.45cm 0.0cm 0.1cm 0.3cm},clip,width=0.9\columnwidth]{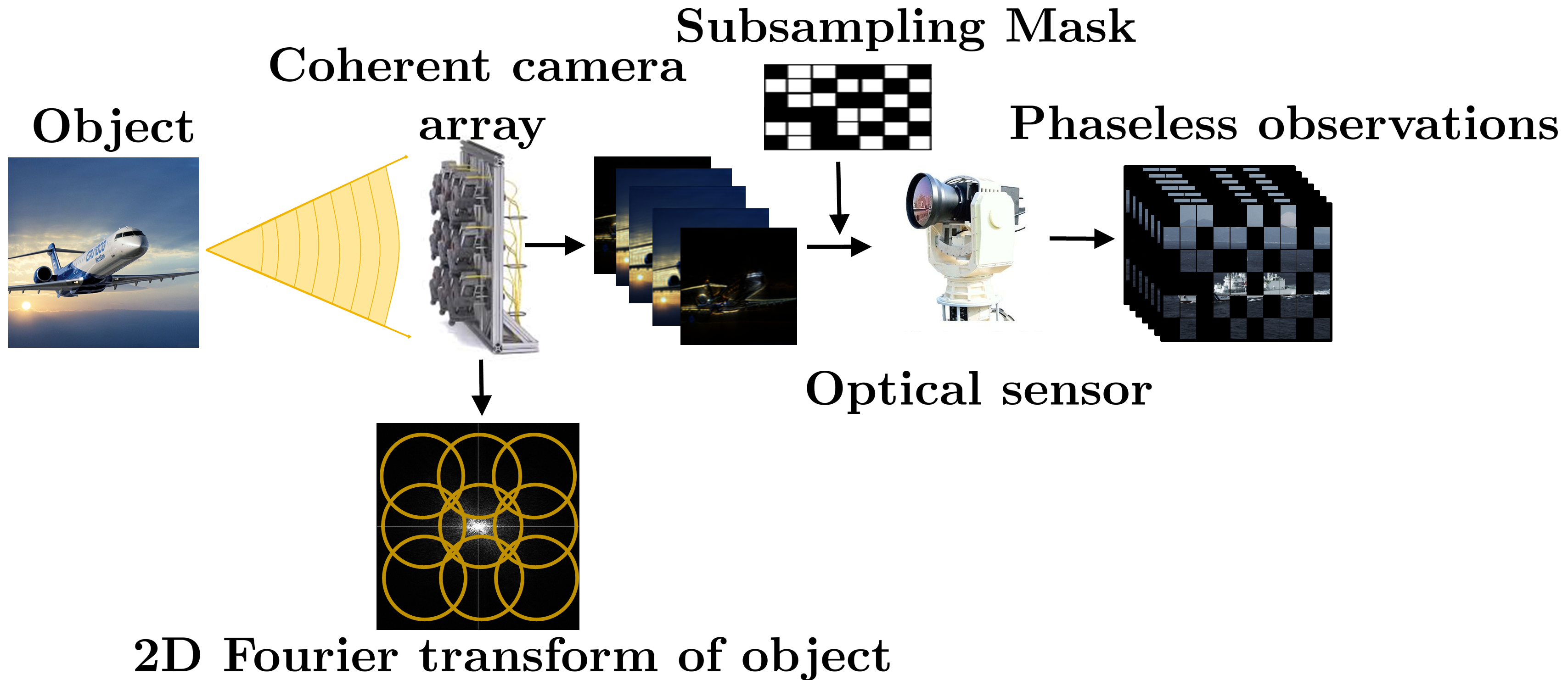}} \hspace{-0.2cm}
\caption{ \small {Fourier Ptychography forward acquistion model. The object has been illuminated using coherent light source. A coherent camera array captures illumination field from the object. The bandlimited signal is then focused to an image plane and a subsampling operator is applied. Subsequently, an optical sensor measures the magnitude while discarding the phase of signal. The effect of atmospheric turbulence causes blurry observations.}}
 \label{fig:fp_setup}
\end{figure}
\begin{figure}[t] \label{fig:blur}
\centering
\raisebox{0.15in}{\rotatebox[origin=t]{90}{\small \hspace{5em} Gaussian}} \hspace{-0.1em}
\subfigure{\includegraphics[width=0.18\columnwidth]{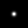}}\hspace{0.0em}
\subfigure{\includegraphics[width=0.18\columnwidth]{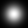}}
\subfigure{\includegraphics[width=0.18\columnwidth]{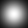}}\hspace{0em}
\subfigure{\includegraphics[width=0.18\columnwidth]{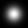}}
\subfigure{\includegraphics[width=0.18\columnwidth]{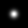}} \\[-2.9em]
\raisebox{0.15in}{\rotatebox[origin=t]{90}{\small \hspace{3.4em} Motion}} \hspace{0.2em}
\subfigure{\includegraphics[width=0.18\columnwidth]{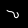}}\hspace{0em}
\subfigure{\includegraphics[width=0.18\columnwidth]{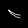}} \hspace{0em}
\subfigure{\includegraphics[width=0.18\columnwidth]{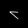}}\hspace{0.0em}
\subfigure{\includegraphics[width=0.18\columnwidth]{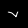}} \hspace{-0.1em}
\subfigure{\includegraphics[width=0.18\columnwidth]{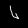}}
\caption{Samples of synthetically generated Gaussian and motion blur kernels.}
\end{figure}
\begin{figure*}[t] \label{fig:qualitative1}
\raisebox{0.15in}{\rotatebox[origin=t]{90}{\small \hspace{4em} Original}} 
\subfigure{\includegraphics[width=0.15\columnwidth]{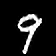}} \hspace{-0.3em}
\subfigure{\includegraphics[width=0.15\columnwidth]{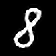}} \hspace{-0.2em}
\subfigure{\includegraphics[width=0.15\columnwidth]{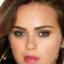}}\hspace{-0.2em}
\subfigure{\includegraphics[width=0.15\columnwidth]{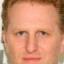}} \hspace{-0.3em}
\subfigure{\includegraphics[width=0.15\columnwidth]{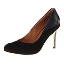}} \hspace{-0.3em}
\subfigure{\includegraphics[width=0.15\columnwidth]{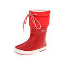}} 
 \\[-2.5em]
 
 \raisebox{0.15in}{\rotatebox[origin=t]{90}{\small \hspace{3em} Range}} 
\subfigure{\includegraphics[width=0.15\columnwidth]{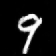}} \hspace{-0.2em}
\subfigure{\includegraphics[width=0.15\columnwidth]{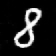}} \hspace{-0.5em}
\subfigure{\includegraphics[width=0.15\columnwidth]{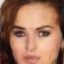}} \hspace{-0.5em} 
\subfigure{\includegraphics[width=0.15\columnwidth]{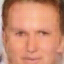}} \hspace{-0.2em}
\subfigure{\includegraphics[width=0.15\columnwidth]{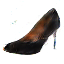}} \hspace{-0.5em} 
\subfigure{\includegraphics[width=0.15\columnwidth]{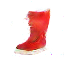}} \hspace{-0.2em}\\[-1.5em]
 
\raisebox{0.15in}{\rotatebox[origin=t]{90}{\small \hspace{3em} 4$\times$}} \hspace{-0.2em}
\subfigure{\includegraphics[width=0.15\columnwidth]{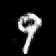}} \hspace{-0.0em}
\subfigure{\includegraphics[width=0.15\columnwidth]{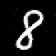}} \hspace{-0.2em}
\subfigure{\includegraphics[width=0.15\columnwidth]{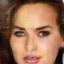}} \hspace{-0.3em}
\subfigure{\includegraphics[width=0.15\columnwidth]{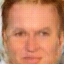}} \hspace{-0.4em}
\subfigure{\includegraphics[width=0.15\columnwidth]{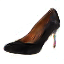}} \hspace{-0.4em}
\subfigure{\includegraphics[width=0.15\columnwidth]{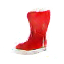}}  \\[-1.5em]

\raisebox{0.15in}{\rotatebox[origin=t]{90}{\small \hspace{3em} 10$\%$}} 
\subfigure{\includegraphics[width=0.15\columnwidth]{./figures/mnist/bdpr_10_subsamp_10RR/im_10_meas_10_RR_1.png}} \hspace{-0.2em}
\subfigure{\includegraphics[width=0.15\columnwidth]{./figures/mnist/bdpr_10_subsamp_10RR/im_10_meas_10_RR_3.png}} 
\subfigure{\includegraphics[width=0.15\columnwidth]{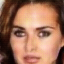}} \hspace{-0.2em}
\subfigure{\includegraphics[width=0.15\columnwidth]{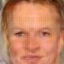}} \hspace{-0.5em}
\subfigure{\includegraphics[width=0.15\columnwidth]{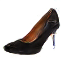}} \hspace{-0.5em}
\subfigure{\includegraphics[width=0.15\columnwidth]{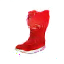}} \hspace{-0.5em}  \\[-1.5em]

\raisebox{0.15in}{\rotatebox[origin=t]{90}{\small \hspace{3em} 1$\%$}} 
\subfigure{\includegraphics[width=0.15\columnwidth]{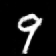}} \hspace{-0.2em}
\subfigure{\includegraphics[width=0.15\columnwidth]{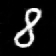}} 
\subfigure{\includegraphics[width=0.15\columnwidth]{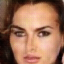}} \hspace{-0.5em}
\subfigure{\includegraphics[width=0.15\columnwidth]{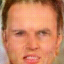}} \hspace{-0.5em}
\subfigure{\includegraphics[width=0.15\columnwidth]{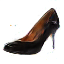}} \hspace{-0.5em}
\subfigure{\includegraphics[width=0.15\columnwidth]{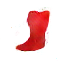}} \hspace{-0.5em}  \\[-0.8em]

\caption{\small  Phaseless blind image deblurring results using generative priors for oversampled Fourier (4x) and subsampled FP measurement (for 1$\%$ and 10$\%$) models for MNIST, CelebA, and Shoes datasets. Deep PBD is able to reconstruct faithful estimates with in the range of the pretrained generative models. Note the conjugate flip in digit 8 (second row) for 4x Fourier measurements due to trivial ambiguities. }
\end{figure*}

\begin{figure} \label{fig:plots}
\centering
\subfigure{\includegraphics[trim={0.4cm 0.4cm 0.4cm 0.58cm},clip,width=0.51\columnwidth]{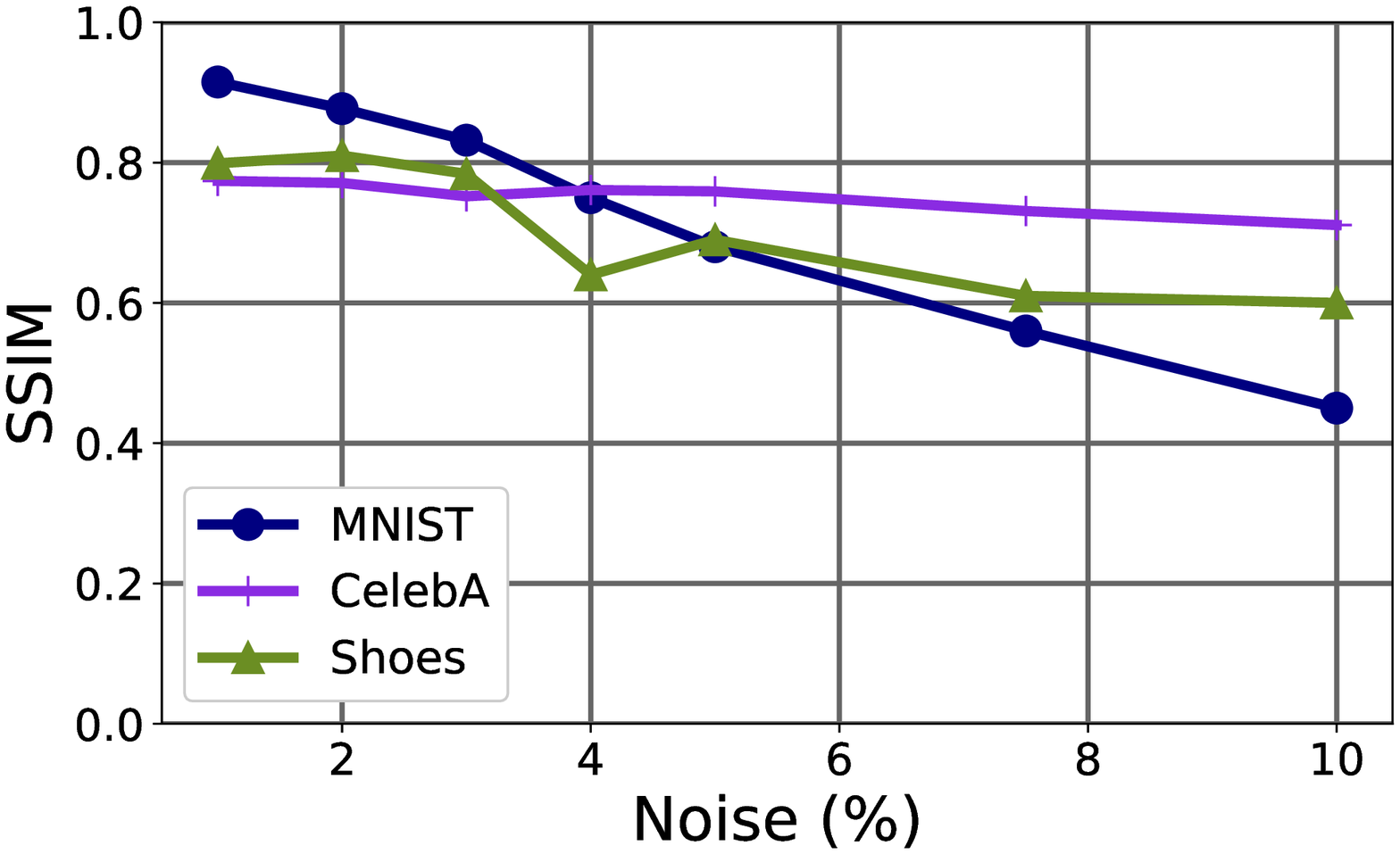}}
\subfigure{\includegraphics[trim={2.4cm 0.4cm 0.4cm 0.65cm},clip,width=0.465\columnwidth]{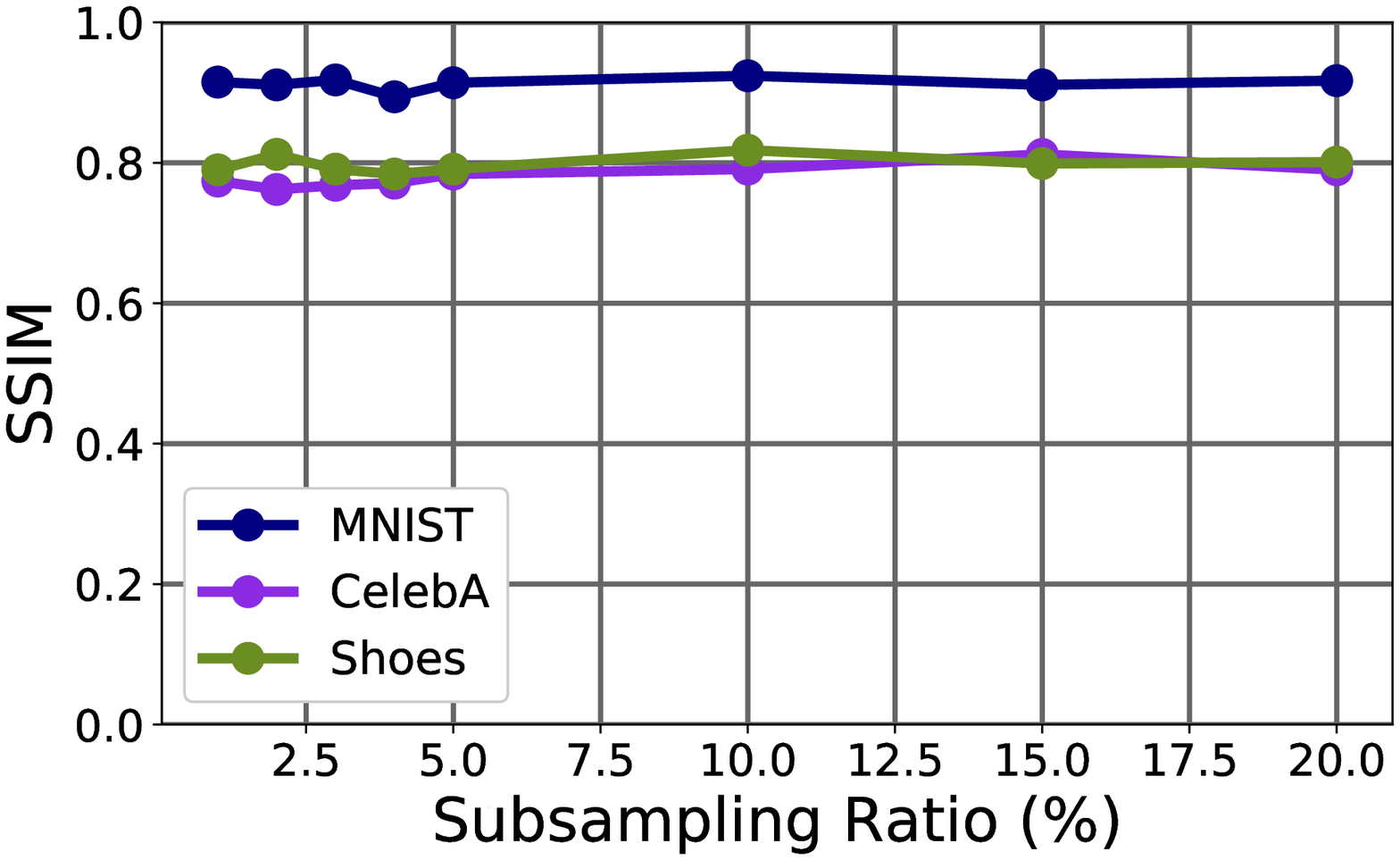}}
\caption{SSIM plots of Deep PBD for subsampled Fourier ptychography model against different noise levels (for 5$\%$ subsampling ratio) and subsampling ratios (for 1$\%$ noise).}
\end{figure}

In this work, we consider two forward operators for evaluating the performance of the proposed approach, Deep PBD. 

\subsection{Fourier Model} In first case, we take forward operator $\mathcal{A}$ as a discrete Fourier Transform (DFT) matrix, denoted by $\mathcal{F}$. The Fourier measurements of real-valued signals are prevalent in many real-world applications including astronomical imaging, imaging through turbulent atmosphere etc. 

\subsection{Subsampled FP Model} In second case, we take forward operator $\mathcal{A}$ as subsampled FP measurement matrix (special case of phase retrieval). FP is an emerging computational imaging technique that shows promising results to mitigate the effects of diffraction blur that is inherited in long-distance imaging \cite{holloway2016toward, shamshad2019deep}. Typical setup of FP is shown in Figure 1. FP works by  capturing illumination field from the object (here $\boldsymbol{i}$) through a coherent camera array (for more details about acquisition model of FP we refer readers to \cite{holloway2016toward}). For $\ell^{th}$ camera of coherent array, the forward operator has the form of $\mathcal{A}_\ell = \mathcal{M}_{\ell}\mathcal{F}^{-1} \mathcal{P}_\ell \circ \mathcal{F}$, where  $\mathcal{F}$ denotes Fourier matrix, $\mathcal{P}_{\ell}$ is a pupil mask that acts as a bandpass filter in the Fourier domain, $\circ$ represents Hadamard product, and $\mathcal{M}_{\ell}$ is subsampling operator. Subsampling operator when applied to measurements $\boldsymbol{y}$, randomly picks a fraction of samples discarding the others \cite{jagatap2018sub}. We define the subsampling ratio as the fraction of samples retained by $\mathcal{M}_\ell$ (for $\ell = 1,2,...,L$) divided by the total number of observed samples i.e. 
\begin{equation}
\text{Subsampling Ratio (\%)} =\frac{\text{Fraction of samples retained ($f$)}\times 100}{\text{Total observed samples ($nL$)}}. \nonumber 
\end{equation}
The subsampling mask resembles the operation of a binary  matrix having entries 1's and 0's. The mask has been element-wise multiplied with the observations in such a way that pixels corresponding to 1's are retained and those corresponding to 0's are discarded. Hence subsampling ratio $f$ governs the percentage of samples that will be retained.

\section{Experiments} \label{sec:experiments}

In this section, we evaluate the performance of Deep PBD both qualitatively and quantitatively. To quantitatively evaluate the performance of our algorithm, we use two metrics, Peak Signal to Noise Ratio (PSNR) and Structural Similarity Index Measure (SSIM). 

\textbf{Image and Blur Datasets}: We evaluate performance of Deep PBD on one grayscale and two RGB image datasets. These datasets include MNIST \cite{deng2012mnist}, CelebA \cite{liu2015deep}, and Shoes \cite{yu2014fine}.  
 For blurring, we use Gaussian and motion blur kernels in experiments. 
Motion blurs having lengths between 5 and 28 are generated following the strategy outlined in \cite{kupyn2018deblurgan}. Gaussian blurs are generated by varying the standard deviation between 0.5 and 1.5.  Visual depictions of both blur datasets is shown in Figure 2. We generate 80,000 blurs for each dataset and split them into 60,000 training and 20,000 test examples.

\textbf{Generator Architecture}: For RGB image datasets, we use the deep convolutional generative adversarial network (DCGAN) \cite{radford2015unsupervised}. DCGAN uses convolutional layers in its generator and discriminator architecture to exploit the hierarchy of representations from image parts. For DCGAN, size of low dimensional latent representation $\boldsymbol{z}$ is set to $100$ and is sampled from a random normal distribution. We train DCGAN model on the training set of low-resolution datasets by updating generator $\mathcal{G}$ twice and discriminator $\mathcal{D}$ once in each cycle to avoid fast convergence of $\mathcal{D}$. Each update during training use the Adam optimizer with batch size 64, $\beta_1 = 0.5$, and learning rate $0.0002$. Generator, after training, is employed as a regularizer for the proposed PBD algorithm.
 For MNIST dataset and blur kernels, we trained VAE with same architecture as proposed in \cite{asim2018solving} having latent dimension of 50. 

\textbf{Experimental Setup}: For all experiments of Deep PBD, we use Adam optimizer for minimizing the loss function with a learning rate of 0.01. We use 20 and 5 random restarts for Fourier and FP measurement models respectively to initialize random latent vectors with 2000 steps per restart and choose reconstruction with minimum measurement error as our final estimate. In all experiments, we use Gaussian blur for Fourier measurement model and motion blur for the subsampled FP measurement model. For all above datasets,  images from test set were sampled and blurry images were produced by convolving with randomly sampled blur kernels from test sets of blur dataset. All our experiments are performed by adding $1\%$\footnote{For an image scaled between 0 and 1, Gaussian noise of $1\%$ translate to Gaussian noise with standard deviation $\sigma = 0.01$ and mean $\mu=0$.} Gaussian noise to phaseless blurry observations $\boldsymbol{y}$, unless stated otherwise. All simulations are performed on core-i7 computer (3.40 GHz and 16GB RAM) equipped with Nvidia Titan X GPU. We use TensorFlow library for implementing the proposed approach.
 \begin{table}[t]
	\caption{\small{Average PSNR (dB) and SSIM values for Fourier measurement model (4x measurements)} and subsampled FP measurement model (with 1$\%$ and 10$\%$ subsampling ratio).}
	\label{table:performance}
	\begin{center}
		
		\resizebox{0.8\textwidth}{!}{
			\begin{tabular}{c|ccc|ccc}
				
				\hline \hline
				
				\multicolumn{5}{c}{\textbf{PSNR}} &  \multicolumn{1}{c}{\textbf{SSIM}}\\

				\textbf{} & \textbf{4x} & \textbf{1 $ \%$}  &  \textbf{10 $ \%$}   & 
				\textbf{4x} & \textbf{1 $ \%$}  &  \textbf{10 $ \%$}   \\
				
				\hline \hline
				\textbf{MNIST}  & 22.50  & 23.85 & 24.32 & 0.909 & 0.915  &0.917  \\
				\textbf{CelebA}  & 21.43  & 21.00 & 22.43 & 0.806 & 0.774  &  0.791 \\
				
				\textbf{Shoes}  & 21.74  & 21.12 & 21.03 & 0.817 & 0.799  & 0.803 \\
				\hline \hline

			\end{tabular} 
		}
	\end{center}
	\vspace{-2em}
\end{table}

\subsection{Qualitative and Quantitative results}

Qualitative results of Deep BPD for oversampled Fourier and subsampled FP forward operators are shown in Figure 3. For Fourier experiments, we oversampled the spectrum by 4 times following the recent work of \cite{metzler2018prdeep}. That is, we first place the  $64 \times 64$ blurry images at the center of $128 \times 128$ square grid and take the 2D Fourier transform of that image. We assumed that the support, i.e., the location of the image within the $128 \times 128$ grid, is known a priori (known as support constraint in phase retrieval literature). As shown in Figure 3, Deep BPD is able to reconstruct quality estimate for phaseless blurry Fourier observations. For subsampled FP model, Deep BPD is able to reconstruct faithful estimates at low subsampling ratios of 1$\%$ and 10$\%$. On a close inspection, it becomes clear that how well the reconstructed output of Deep PBD approximates the true image roughly depends on how close the corresponding range image is to true image exactly. This observation is consistent with the other works related to solving inverse problems using generative models \cite{bora2017compressed}, Fourier ptychography \cite{shamshad2018robust}, phase retrieval \cite{hand2018phase}. 

Quantitative results, in terms of PSNR and SSIM, for MNIST, CelebA, and Shoes dataset are shown in Table \ref{table:performance}. The results are
averaged over randomly selected 20 images from the test set of each dataset. In the case of Fourier measurements, we take care of the trivial ambiguities (reflections and translations) before evaluating PSNR and SSIM values. From Table \ref{table:performance}, it can be seen that Deep PBD is able to achieve reasonable PSNR and SSIM values. In Figure 4, we show SSIM plots of Deep PBD for subsampled Fourier ptychography model against different noise levels (for 5$\%$ subsampling ratio) and subsampling ratios (for 1$\%$ noise). It can be seen that the proposed approach is robust to high additive noise, especially for GAN based pre-trained generative models, for subsampled FP forward operator. Further, the Deep PDB (for FP forward operator) is able to achieve reasonable performance (within the range of the generative model) at very low subsampling ratios as can be visualized in Figure 4 (right plot).

\section{Conclusion and Future Directions} \label{sec:conclusion}
To conclude, we demonstrate the effectiveness of integrating deep generative priors with the relatively unexplored and ill-posed problem of phaseless blind image deblurring. Our preliminary results indicate that generative priors can effectively regularize the otherwise ill-posed phaseless blind image deblurring problem. We observe that the proposed approach struggles for Fourier measurements and performance depends heavily on the number of random restarts and size of the blur kernel. One way to circumvent this issue is to propose an effective initialization strategy, as in \cite{metzler2018prdeep}. Further, the output of the proposed approach is constrained to lie in the range of the pre-trained generative model. We can mitigate this range by using invertible generative models as priors that have zero representation error by design \cite{asim2019invertible, shamshad2019subsampled}. Further, we will aim to extend the proposed approach to natural images by leveraging the prior imposed by the structure of the untrained generative model \cite{ulyanov2018deep}. This allows us to extend our deblurring algorithm from compact image datasets such as face images to more complex, larger images such as natural scenes that require a heavy compute resource to reliably train the generative network. We leave these directions and extensive experimental analysis of Deep PBD as our future work.

\bibliographystyle{unsrt}
\bibliography{refs}
\end{document}